\documentclass[pre,twocolumn,aps,floatfix,10pt]{revtex4}
\usepackage{amsmath}
\usepackage{amsfonts}
\usepackage{amssymb}
\usepackage{graphicx}
\usepackage{fontenc}
\usepackage{psfrag,layout}
\usepackage{upgreek}

\begin{document}
\title{Effective Diffusion and transport coherence in presence of inhomogeneous temeprature: Piecewise linear potential}
\author{Ronald Benjamin}
\affiliation{Institut f{\"u}r Theoretische Physik II, Universit{\"a}t D{\"u}sseldorf,
Universit\"atsstra\ss e 1, 40225 D{\"u}sseldorf, Germany}

\begin{abstract}
We compute the effective diffusion coefficient of a Brownian particle
in a piece-wise linear periodic potential and subject of spatially inhomogeneous
temperature, otherwise known as the B{\"u}ttiker-Landauer motor. We obtain analytical
expressions for the current and diffusion coefficients and compare with numerical results.

\end{abstract}

\maketitle

\section{Introduction}

A Brownian particle moving in a periodic potential and subected
to a spatially non-uniform temperature profile gives rise
to a net current, acting like a Brownian motor. This device
is autonomous since it is entirely driven by thermal fluctations.
Several properties of such a Brownian motor, such as current,
heat and thermodynamic eficiency have been studied. 

Another important performance characteristic of such a Brownian
device is the transport coherence as measured by the Peclet 
number, which is the ratio of the thermal velocity times the
period of the substrate potential and the effective diffusion
coefficient of the motor. It is desirable to design motors
which produce the maximum velocity with the minimum dispersion
i.e. small diffusion coefficient. 

While the net current of such a Brownian ratchet has been derived
for an overdamped system in the works of Landauer, Van Kampen and
B{\"u}ttiker the determination of the effective diffusion coefficient had
remained a challenging task until the early twenty-first century.
Reimann {\it{et al.}} first determined the effective diffusion coefficient
for 

In the past decade several studies have appeared regarding the 
coherent transport of Brownian motors. Most studies have been carried out based on
uniform temperature. In this work we obtain analytical expressions for current, effective diffusion 
coefficient analytically and numerically. In the low temperature regime, we determine
the transition rates and from that the current and effective diffusion coefficient.

\section{System}

The potential and temperature profile are respectively:-
 
 $$
  U(x) =
  \begin{cases}
    \frac{U_{0}x}{\alpha L}, & \text{for } 0 \leq x < \alpha L \\
    \newline\\
    \frac{U_{0}(L-x)}{(1-\alpha)L} & \text{for } \alpha L \leq x < L \\
         
  \end{cases}
$$

$$
  T(x) =
  \begin{cases}
    T_H, & \text{for } 0 \leq x < \alpha L \\
    \newline\\
    T_C & \text{for } \alpha L \leq x < L \\
         
  \end{cases}
$$
Both Potential and Temperatue profiles are periodic i.e. $U(x+L)=U(x)$ and $T(x+L)=T(x)$.

\begin{figure}[tbh]
\includegraphics[width=3.0in]{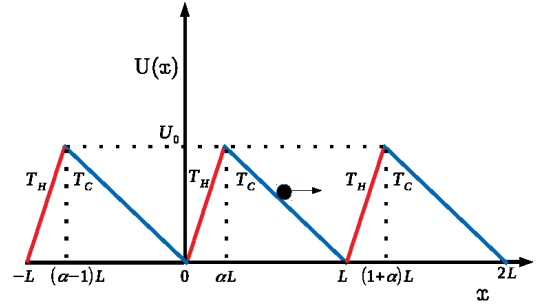}
\caption{\label{fig:profile}(Color online)Schematic of piecewise linear potential alternately sujected to hot and cold baths. }
\end{figure}

$\alpha$ is the potential asymmetry parameter, $U_{0}$ is the barrier height and $L$ is the spatial period of the potential. Without loss of generality we consider $U_{0}=1$ and $L=1$.

The Langvin equation used to study the motion of the Brownian particle is
given by,
\begin{equation}
 m\ddot{x}=-\gamma\dot{x}-U^\prime(x)+\sqrt{2k_{B}T(x)\gamma}\xi(t)
\end{equation}
We set $k_{B}=1$ and $\gamma=1$.

\begin{figure}
 \includegraphics[width=3.0in]{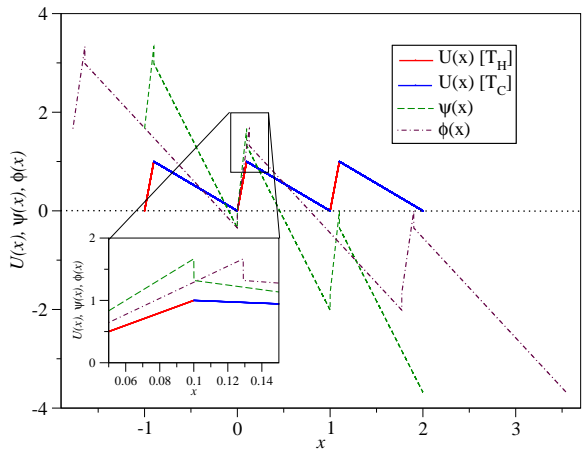}
 \caption{\label{fig:potential}(Color online) $U(x)$, $\psi(x)$, and $\phi(x)$.}
 \end{figure}

In the overdamped limit, we ignore the inertial term. However, for temperature dependent on position, an additional term needs to be added as pointd out in earlier works. As per the Stratonovich interpretation, the overdamped Langevin equation is given by,

\begin{equation}
  \dot{x}=-U^\prime(x) -\frac{1}{2}\frac{dT(x)}{dx} + \sqrt{2T(x)\gamma}\xi(t)
\end{equation}
Here, $<\xi(t)>=0$ and $<\xi(t)\xi(t^\prime)>=\delta(t-t^\prime)$.
We set $\gamma=1$. This equation will be understood according to the Stratonovivh
interpretation.

The Fokker-Planck equation corresponding to this overdamped Langevin equation is given by,

\begin{equation}
 \frac{\partial P(x,t)}{\partial t}=\frac{\partial }{\partial x}\left[U^\prime(x)P(x,t)\right]+
 \frac{\partial^{2}}{\partial x^{2}}\left[T(x) P(x,t)\right]
\end{equation}

The current  is given by,
\begin{equation}
 <\dot{x}>=\frac{x(t_f)-x(t_s)}{t_f - t_s}
\end{equation}
where, $t_s \gg 1$ is the time taken to reach the steady state and $t_f$ 
is the final time upto which the simulations are carried out.

The effective diffusion coefficient is computed as per the following relation:
\begin{equation}
 D_{eff}=\frac{<(x(t_f)-x(t_s))^{2}>-<x(t_f)-x(t_s)>^2}{2(t_f-t_s)}
\end{equation}

Our numerical calculations are carrier out as per the Stochastic 
Euler-Maruyama algorithm.

The analytical calculation were carried out as per the following formulas 
provided in Ref.~\cite{lindner}.

\begin{equation}
 <\dot{x}>=L\frac{1-\exp(\psi(L))}{\int_{0}^{L} dx I_+(x)/g(x)}
\end{equation}
where, $g(x)=\sqrt{T(x)}$ and 
\begin{equation}
 I_+(x)=\exp(-\psi(x))\int_{x}^{x+L} dy \exp(\psi(y))/g(y)
\end{equation}

The analytical expression for the effective diffusion coefficient is,
\begin{equation}
 D_{eff}=(L^{2})\frac{\int_{0}^{L} dx I_+(x)^{2}I_-(x)/g(x)}{[\int_{0}^{L} dx I_+(x)/g(x)]^{3}}
\end{equation}
where,
\begin{equation}
 I_-(x)=\exp(\psi(x))\int_{x-L}^{x} dy \exp(-\psi(y))/g(y)
\end{equation}
We can also determine the Peclet number which determines the coherence of transport. It's
given by,
\begin{equation}
 Pe=L <\dot{x}>/D_{eff}
\end{equation}

We also carried out numerical simulations to test the validity of our analytical 
results using the Euler-Mayuram algorithm. The time-step chosen was $h=0.001$ and
he number of relizations was $5000$.

\section{Exact Expressions}
The first step to solve this model is to recognize that the 
non-uniform temperature breaks the symmetry and simultaneously results in
a non-equilibriun condition, the minimal ingredient to 
produce directed motion. In order  break the left right symmetry there should 
be a phase difference between them. Due to the spatial dependence of the temperature
profile the noise term is multiplicative and Brownian particles subject
to such a temperature profile move under the influence of the generalized
potential given by,
\begin{equation}
 \psi(x)=\int_{0}^{x} dx^\prime [U^{\prime}(x^{\prime})+(1/2)T^{\prime}(x^{\prime})]/T(x^{\prime})
\end{equation}
The condition to achieve  directed transport ($ <\dot{x}>\neq 0$) is for this potential to have an
effective bias such that $\psi(L=1)-\psi(0) \neq 0$.

For the piecewise linear potential  and piecewise constant temperature profile,
it is simple to calculate $\psi(x)$, which is given by,
\begin{widetext}
 $$
  \psi(x) =
  \begin{cases}
    \frac{U_0}{T_C}+\frac{U_{0}(x+1-\alpha)}{\alpha T_H}, & \text{for } -1 \leq x < -1+\alpha  \\
    \newline\\
    \frac{-U_{0}x}{(1-\alpha)T_C} +\frac{1}{2}\log(\frac{T_C}{T_H}) & \text{for } -1+\alpha \leq x < 0 \\
    \newline \\
     \frac{U_{0}x}{\alpha T_H}    & \text{for } 0 \leq x <\alpha \\
     \newline \\
     \frac{U_0}{T_H}-\frac{U_{0}(x-\alpha)}{(1-\alpha)T_{C}}+\frac{1}{2}\log(\frac{T_C}{T_H})  & \text{for }  \alpha \leq x < 1 \\
     \newline \\
     \frac{U_0}{T_H}-\frac{U_0}{T_C}+\frac{U_0(x-1)}{\alpha T_H} & \text{for }  1 \leq x < 1+\alpha \\
     \newline \\
     \frac{2U_0}{T_H}-\frac{U_0}{T_C}-\frac{U_{0}(x-1-\alpha)}{(1-\alpha)T_C}+\frac{1}{2}\log(\frac{T_C}{T_H})  &    \text{for }  1+\alpha \leq x < 2 \\
  \end{cases} 
$$

\end{widetext}

Using a suitable transformation of variables one can convert the overdamped LAngevin equation with multiplicatie noise to one with additive noise. 
The required transformation is given by,
\begin{equation}
 y(x)=\int_{0}^{x} \,dz/\sqrt{T(z)}
\end{equation}
and the corresponding Langevin equation is given by,

\begin{equation}
 \dot{y}=\dot{x}/\sqrt{T(x)}=-\frac{d\phi(y)}{dy}+\sqrt{2}\xi(t)
\end{equation}
where,
\begin{equation}
 \phi(y)=\int_{0}^{y} \,dy^{*} \frac{U^{\prime}[x(y^{*})]+(1/2)T^{\prime}[x(y^{*})]}{\sqrt{T[x(y^{*})]}}
\end{equation}
and  $\phi(y)=\psi[x(y)]\,.$

For our potential and temperature profiles, the relation between the original and transformed coordinates is given by,

 $$
  y =
  \begin{cases}
    \frac{\alpha-1}{\sqrt{T_C}}-\frac{\alpha-x-1}{\sqrt{T_H}}, & \text{for } -1 \leq x < -1+\alpha  \\
    \newline\\
    \frac{x}{\sqrt{T_C}}& \text{for } -1+\alpha \leq x < 0 \\
    \newline \\
     \frac{x}{\sqrt{T_H}}    & \text{for } 0 \leq x <\alpha \\
     \newline \\
     \frac{1}{\sqrt{T_H}}+\frac{x-\alpha}{\sqrt{T_{C}}}  & \text{for }  \alpha \leq x < 1 \\
     \newline \\
     \frac{1}{\sqrt{T_H}}+\frac{1-\alpha}{\sqrt{T_C}}+\frac{(x-1)}{\sqrt{T_H}} & \text{for }  1 \leq x < 1+\alpha \\
     \newline \\
     \frac{2\alpha}{\sqrt{T_H}}+\frac{1-\alpha}{\sqrt{T_C}}+\frac{(x-1-\alpha)}{\sqrt{T_C}}  &    \text{for }  1+\alpha \leq x < 2 \\
  \end{cases} 
$$

Finally, the effective potential in the transformed coordinates is given by,

 \begin{widetext}
 $$
  \phi(y) =
  \begin{cases}
    \frac{U_0}{T_C}+\frac{U_{0}(y+[(1-\alpha)/\sqrt{T_C}])}{\alpha \sqrt{T_H}}, & \text{for } \frac{\alpha-1}{\sqrt{T_C}}-\frac{\alpha}{\sqrt{T_H}} \leq y < \frac{\alpha-1}{\sqrt{T_C}}  \\
    \newline\\
    \frac{-U_{0}y}{(1-\alpha)\sqrt{T_C}} +\frac{1}{2}\log(\frac{T_C}{T_H}) & \text{for } \frac{-1+\alpha}{\sqrt{T_C}} \leq y < 0 \\
    \newline \\
     \frac{U_{0}y}{\alpha \sqrt{T_H}}    & \text{for } 0 \leq y < \frac{\alpha}{\sqrt{T_H}} \\
     \newline \\
     \frac{U_0}{T_H}-\frac{U_{0}(y-\alpha/\sqrt{T_H})}{(1-\alpha)\sqrt{T_{C}}}+\frac{1}{2}\log(\frac{T_C}{T_H})  & \text{for }  \frac{\alpha}{\sqrt{T_H}} \leq y < \frac{\alpha}{\sqrt{T_H}} + \frac{1-\alpha}{\sqrt{T_C}} \\
     \newline \\
     -\frac{U_0}{T_C}+\frac{U_0(y+(\alpha-1)/\sqrt{T_C})}{\alpha \sqrt{T_H}} & \text{for }  \frac{\alpha}{\sqrt{T_H}} + \frac{1-\alpha}{\sqrt{T_C}} \leq y < \frac{2\alpha}{\sqrt{T_H}} + \frac{1-\alpha}{\sqrt{T_C}} \\
     \newline \\
     \frac{2U_0}{T_H}-\frac{U_0}{T_C}-\frac{U_{0}(y-2\alpha/\sqrt{T_H})}{(1-\alpha)\sqrt{T_C}}+\frac{1}{2}\log(\frac{T_C}{T_H})  &    \text{for }  \frac{2\alpha}{\sqrt{T_H}} + \frac{1-\alpha}{\sqrt{T_C}}  \leq y < \frac{2\alpha}{\sqrt{T_H}} + \frac{2(1-\alpha)}{\sqrt{T_C}}  \\
  \end{cases} 
$$

\end{widetext}
The period of $\phi(y)$ is $L_{y}=\frac{\alpha}{\sqrt{T_H}}+\frac{1-\alpha}{\sqrt{T_C}}$.

The effective diffusion coefficient computed as per the transformed dynamics with additive noise
is given by,
\begin{equation}
 D_{eff,y}=\frac{\int_{0}^{L_y} \,dx [I_-(x)]^{2}I_+(x)/L_y}{[\int_{0}^{L_y} \,dx I_-(x)/L_y]^{3}}
\end{equation}
and
\begin{equation}
 I_\pm (x)=\mp e^{\pm \phi(x)}\int_{x}^{x \mp L_y} \,dy \, e^{\mp \phi(y)}
\end{equation}

The velocity in the transformed coordinates is given by,
\begin{equation}
 v_{y}=\frac{1-e^{\phi(L_y)}}{[\int_{0}^{L_y} \,dx \, I_-(x)/L_y]}
\end{equation}

The relation between the effective diffusion coefficient and the particle current
in the original and transformed coordinates is given by,
\begin{equation}
 D_{eff}=\frac{D_{eff,y}}{{L_{y}}^{2}} , \,\,\, <\dot{x}>=\frac{v_{y}}{L_{y}}
 \end{equation}

 We will calculate the effective diffusion coefficient using Eq. 8.
 The Integral in the denominatior is given by,
 \begin{widetext}
 \begin{equation}
  I_d=\int_{0}^{\alpha} \,dx \frac{e^{-\psi(x)}}{\sqrt{T_H}} \int_{x}^{x+1} \,dy \frac{e^{\psi(y)}}{\sqrt{T(y)}} +   \int_{\alpha}^{1} \,dx \frac{e^{-\psi(x)}}{\sqrt{T_C}} \int_{x}^{x+1} \,dy \frac{e^{\psi(y)}}{\sqrt{T(y)}}= A+B
 \end{equation}
\end{widetext}
such that 
\begin{equation}
 A=\int_{0}^{\alpha} \,dx \frac{e^{-\psi(x)}}{\sqrt{T_H}} \int_{x}^{x+1} \,dy e^{\psi(y)}/\sqrt{T(y)}
\end{equation}
and,
\begin{equation}
 B=\int_{\alpha}^{1} \,dx \frac{e^{-\psi(x)}}{\sqrt{T_C}} \int_{x}^{x+1} \,dy e^{\psi(y)}/\sqrt{T(y)}
\end{equation}
we find that,
\begin{widetext}
\begin{equation}
 A=\int_{0}^{\alpha} \,dx \frac{e^{-\frac{U_0{x}}{\alpha T_H}}}{\sqrt{T_H}}\left[ \int_{x}^{\alpha} \,dy \frac{e^{\frac{U_0 {y}}{\alpha T_H}}}{\sqrt{T_H}} + \int_{\alpha}^{1} \,dy \frac{e^{\frac{U_0}{T_H}-\frac{U_{0}(y-\alpha)}{(1-\alpha)T_{C}}+\frac{1}{2}\log\left(\frac{T_C}{T_H}\right) }}{\sqrt{T_C}} + \int_{1}^{x+1} \,dy \frac{e^{\frac{U_0}{T_H}-\frac{U_0}{T_C}+\frac{U_0(y-1)}{\alpha T_H}}}{\sqrt{T_H}} \right]
\end{equation}
and,
\begin{equation}
\begin{split}
 B   = \int_{\alpha}^{1} \,dx \frac{e^{-\left[{\frac{U_0}{T_H}-\frac{U_{0}(x-\alpha)}{(1-\alpha)T_{C}}+\frac{1}{2}\log(\frac{T_C}{T_H})  }\right]}}{\sqrt{T_C}} \left[ \int_{x}^{1}  \,dy \frac{e^{\frac{U_0}{T_H}-\frac{U_{0}(y-\alpha)}{(1-\alpha)T_{C}}+\frac{1}{2}\log\left(\frac{T_C}{T_H}\right) }}{\sqrt{T_C}} +\int_{1}^{1+\alpha} \,dy \frac{e^{\frac{U_0}{T_H}-\frac{U_0}{T_C}+\frac{U_0(y-1)}{\alpha T_H}}}{\sqrt{T_H}} \right.  \\
 \left.   +\int_{1+\alpha}^{x+1}\,dy \frac{e^{\frac{2U_0}{T_H}-\frac{U_0}{T_C}-\frac{U_{0}(y-1-\alpha)}{(1-\alpha)T_C}+\frac{1}{2}\log(\frac{T_C}{T_H})}} {\sqrt{T_C}}  \right]
\end{split}    
 \end{equation}

\end{widetext}
Here,
\begin{equation}
 a_{11}=\int_{x}^{\alpha} \,dy  \frac{e^{\frac{U_0 {y}}{\alpha T_H}}}{\sqrt{T_H}}=-{\frac {\alpha\,\sqrt {T_{H}}}{U_{0}} \left( {{\rm e}^{{\frac {U_{0}
\,x}{\alpha\,T_{H}}}}}-{{\rm e}^{{\frac {U_{0}}{T_{H}}}}} \right) }
\end{equation}
\begin{widetext}
\begin{equation}
 a_{12}=\int_{\alpha}^{1} \,dy  \frac{e^{\frac{U_0}{T_H}-\frac{U_{0}(y-\alpha)}{(1-\alpha)T_{C}}+\frac{1}{2}\log\left(\frac{T_C}{T_H}\right) }}{\sqrt{T_C}} = 
 -{\frac { \left( \alpha-1 \right) T_{C}}{U_{0}\,\sqrt {T_{H}}}{{\rm e}
^{{\frac {U_{0}\, \left( T_{C}-T_{H} \right) }{T_{C}\,T_{H}}}}}
 \left( {{\rm e}^{{\frac {U_{0}}{T_{C}}}}}-1 \right) }
\end{equation}
\end{widetext}
and,
\begin{widetext}
\begin{equation}
 a_{13}=\int_{1}^{x+1} \frac{e^{\frac{U_0}{T_H} -\frac{U_0}{T_C} + \frac{U_{0}(y-1)}{\alpha T_H} }   } {\sqrt{T_H}} = {\frac {\alpha\,\sqrt {T_{H}}}{U_{0}} \left( {{\rm e}^{{\frac {U_{0}\,
 \left( \alpha+x \right) }{\alpha\,T_{H}}}}}-{{\rm e}^{{\frac {U_{0}}{
T_{H}}}}} \right) {{\rm e}^{-{\frac {U_{0}}{T_{C}}}}}}
\end{equation}
\end{widetext}
\begin{widetext}
Similarly,
\begin{equation}
 b_{11}={\frac { \left( \alpha-1 \right) T_{C}}{U_{0}\,\sqrt {T_{H}}} \left( -
{{\rm e}^{{\frac {U_{0}\, \left( T_{C}\,\alpha+T_{H}\,x-T_{C}-T_{H}
 \right) }{T_{H}\, \left( \alpha-1 \right) T_{C}}}}}+{{\rm e}^{{\frac 
{U_{0}}{T_{H}}}}} \right) {{\rm e}^{-{\frac {U_{0}}{T_{C}}}}}}
\end{equation}
\begin{equation}
 b_{12}={\frac {\alpha\,\sqrt {T_{H}}}{U_{0}}{{\rm e}^{{\frac {U_{0}\, \left( 
T_{C}-T_{H} \right) }{T_{C}\,T_{H}}}}} \left( {{\rm e}^{{\frac {U_{0}
}{T_{H}}}}}-1 \right) }
\end{equation}
\begin{equation}
 b_{13}={\frac { \left( \alpha-1 \right) T_{C}}{U_{0}\,\sqrt {T_{H}}} \left( {
{\rm e}^{{\frac {U_{0}\, \left( 2\,T_{C}\,\alpha-\alpha\,T_{H}+T_{H}\,
x-2\,T_{C} \right) }{T_{H}\, \left( \alpha-1 \right) T_{C}}}}}-{
{\rm e}^{2\,{\frac {U_{0}}{T_{H}}}}} \right) {{\rm e}^{-{\frac {U_{0}
}{T_{C}}}}}}
\end{equation}

\begin{equation}
\begin{split}
 A=-\frac {\alpha}{{U_{0}}^{2}} \left(  \left(  \left( -T_{C}+T_{H}-U_{0
} \right) \alpha+T_{C} \right) {{\rm e}^{{\frac {U_{0}\, \left( T_{C}-
T_{H} \right) }{T_{C}\,T_{H}}}}}+ \right.  \\
\left. \left(  \left( T_{C}-T_{H} \right) 
\alpha-T_{C} \right) {{\rm e}^{-{\frac {U_{0}}{T_{C}}}}}+
\left( 
 \left( T_{C}-T_{H} \right) \alpha-T_{C} \right) {{\rm e}^{{\frac {U_{
0}}{T_{H}}}}}+ \left( -T_{C}+T_{H}+U_{0} \right) \alpha+T_{C} \right) 
\end{split}
\end{equation}

\begin{equation}
 \begin{split}
 B=\frac {\alpha-1}{{U_{0}}^{2}} \left(  \left(  \left( -T_{C}+T_{H}-U_{
0} \right) \alpha+T_{C}+U_{0} \right) {{\rm e}^{{\frac {U_{0}\,
 \left( T_{C}-T_{H} \right) }{T_{C}\,T_{H}}}}}+ \right. \\
 \left. \left(  \left( T_{C}-T
_{H} \right) \alpha-T_{C} \right) {{\rm e}^{-{\frac {U_{0}}{T_{C}}}}}+
 \left(  \left( T_{C}-T_{H} \right) \alpha-T_{C} \right) {{\rm e}^{{
\frac {U_{0}}{T_{H}}}}}+ \left( -T_{C}+T_{H}+U_{0} \right) \alpha+T_{C
}-U_{0} \right) 
 \end{split}
\end{equation}
Then $I_d$ is given by,
\begin{equation}
\begin{split}
 I_d=\frac {1}{{U_{0}}^{2}} \left(  \left(  \left( T_{C}-T_{H}+2\,U_{0}
 \right) \alpha-T_{C}-U_{0} \right) {{\rm e}^{{\frac {U_{0}\, \left( T
_{C}-T_{H} \right) }{T_{C}\,T_{H}}}}}+ \right. \\
\left. \left(  \left( -T_{C}+T_{H}
 \right) \alpha+T_{C} \right) {{\rm e}^{-{\frac {U_{0}}{T_{C}}}}}+
 \left(  \left( -T_{C}+T_{H} \right) \alpha+T_{C} \right) {{\rm e}^{{
\frac {U_{0}}{T_{H}}}}}+ \left( T_{C}-T_{H}-2\,U_{0} \right) \alpha-T_
{C}+U_{0} \right) 
\end{split}
\end{equation}

\end{widetext}

The numerator in the expression for effective diffusion coefficient can be written as,
\begin{widetext}
\begin{equation}
\begin{split}
 num=\int_{0}^{\alpha} \,dx \frac{e^{-\frac{U_0{x}}{\alpha T_H}}}{\sqrt{T_H}} \left[\int_{x}^{x+L}\,dy \frac{\exp(\psi(y))}{\sqrt{T(y)}}\right]^{2}\left(\int_{x-L}^{x}\,dy \frac{\exp[-\psi(y)]}{\sqrt{T(y)}}\right) + \\
 \int_{\alpha}^{1} \,dx  \frac{e^{-\left[{\frac{U_0}{T_H}-\frac{U_{0}(x-\alpha)}{(1-\alpha)T_{C}}+\frac{1}{2}\log(\frac{T_C}{T_H})  }\right]}}{\sqrt{T_C}}\left[\int_{x}^{x+L}\,dy \frac{\exp(\psi(y))}{\sqrt{T(y)}}\right]^{2}\left(\int_{x-L}^{x}\,dy \frac{\exp[-\psi(y)]}{\sqrt{T(y)}}\right)
 \end{split}
\end{equation}
such that,
\begin{equation}
\begin{split}
  num=\int_{0}^{\alpha} \,dx \frac{e^{-\frac{U_0{x}}{\alpha T_H}}}{\sqrt{T_H}} \left[ a_{11}+a_{12}+a_{13} \right]^{2}\left(\int_{x-L}^{x}\,dy \frac{\exp[-\psi(y)]}{\sqrt{T(y)}}\right) + \\
 \int_{\alpha}^{1} \,dx  \frac{e^{-\left[{\frac{U_0}{T_H}-\frac{U_{0}(x-\alpha)}{(1-\alpha)T_{C}}+\frac{1}{2}\log(\frac{T_C}{T_H})  }\right]}}{\sqrt{T_C}}\left[ b_{11}+b_{12}+b_{13}\right]^{2}\left(\int_{x-L}^{x}\,dy \frac{\exp[-\psi(y)]}{\sqrt{T(y)}}\right)
 \end{split}
 \end{equation}
\end{widetext}
For $0 \leq x < \alpha$,
\begin{equation}
 P=\int_{x-1}^{x} \,dy \frac{\exp[-\psi(y)]}{\sqrt{T(y)}}=p_{11}+p_{12}+p_{13}
\end{equation}
where,
\begin{equation}
 p_{11}=
 {\frac {\alpha\,\sqrt {T_{H}}}{U_{0}} \left( {{\rm e}^{{\frac {U_{0}\,
 \left( \alpha-x \right) }{\alpha\,T_{H}}}}}-1 \right) {{\rm e}^{-{
\frac {U_{0}}{T_{C}}}}}}
%
\end{equation}
\begin{equation}
 p_{12}=-{\frac {\sqrt {T_{H}} \left( \alpha-1 \right) }{U_{0}} \left( {
{\rm e}^{{\frac {U_{0}}{T_{C}}}}}-1 \right) {{\rm e}^{-{\frac {U_{0}}{
T_{C}}}}}}
\end{equation}
\begin{equation}
 p_{13}=-{\frac {\alpha\,\sqrt {T_{H}}}{U_{0}} \left( {{\rm e}^{-{\frac {U_{0}
\,x}{\alpha\,T_{H}}}}}-1 \right) }
\end{equation}

For $\alpha \leq x < 1$, 
\begin{equation}
 Q=\int_{x-1}^{x} \,dy \frac{\exp[-\psi(y)]}{\sqrt{T(y)}}=q_{11}+q_{12}+q_{13}
\end{equation}
where,
\begin{widetext}
\begin{equation}
 q_{11}=\int_{x-1}^{0} \,dy \frac{e^{\frac{U_{0}y}{(1-\alpha)T_C}-\frac{1}{2}\log(T_{C}/T_{H})}}{\sqrt{T_C}}=
 {\frac { \left( \alpha-1 \right) \sqrt {T_{H}}}{U_{0}} \left( {{\rm e}
^{-{\frac {U_{0}\, \left( x-1 \right) }{ \left( \alpha-1 \right) T_{C}
}}}}-1 \right) }
%
\end{equation}
\end{widetext}
\begin{equation}
 q12=\int_{0}^{\alpha} \,dy \frac{e^{\frac{-U_{0}y}{\alpha T_H}}}{\sqrt{T_H}}= 
 {\frac {\alpha\,\sqrt {T_{H}}}{U_{0}} \left( {{\rm e}^{{\frac {U_{0}}{
T_{H}}}}}-1 \right) {{\rm e}^{-{\frac {U_{0}}{T_{H}}}}}}
%
\end{equation}

\begin{equation}
\begin{split}
 q13=\int_{\alpha}^{x}\,dy \frac {e^{\frac{-U_{0}}{T_H}+\frac{U_{0}(y-\alpha)}{(1-\alpha)T_{C}}-\frac{1}{2}\log(\frac{T_C}{T_H})} } {\sqrt{T_C}}
 =\\
 -{\frac { \left( \alpha-1 \right) \sqrt {T_{H}}}{U_{0}} \left( {
{\rm e}^{{\frac {U_{0}\, \left( \alpha-x \right) }{ \left( \alpha-1
 \right) T_{C}}}}}-1 \right) {{\rm e}^{-{\frac {U_{0}}{T_{H}}}}}}
\end{split}
\end{equation}

Finally, we have 
\begin{widetext}
\begin{equation}
\begin{split}
num=\int_{0}^{\alpha} \,dx \frac{e^{-\frac{U_0{x}}{\alpha T_H}}}{\sqrt{T_H}} \left[ a_{11}+a_{12}+a_{13} \right]^{2}( p_{11}+p_{12}+p_{13} ) + \\
\int_{\alpha}^{1} \,dx  \frac{e^{-\left[{\frac{U_0}{T_H}-\frac{U_{0}(x-\alpha)}{(1-\alpha)T_{C}}+\frac{1}{2}\log(\frac{T_C}{T_H})  }\right]}}{\sqrt{T_C}}\left[ b_{11}+b_{12}+b_{13}\right]^{2} ( q_{11}+q_{12}+q_{13})
\end{split}
\end{equation}
\end{widetext}
The final expression is obtained as,
\begin{equation}
 num=num1+num2
\end{equation}
where,
\begin{widetext}
\begin{equation}
\begin{split}
 num1=\frac{1}{\sqrt{T_H}}\left({\varphi_{0}}^{3}\alpha +\frac{\alpha T_H}{U_0}\left(\exp\left(\frac{U_{0}}{T_H}\right)-1\right)\left[ {\varphi_{0}}^{2}{\tilde{\varphi_{1}}}{\varphi_{c}}+\varphi_{1}\varphi_{c}(2{\varphi_{0}}^{2}   +\varphi_{1} {\tilde{\varphi_{1}}} {{\varphi_{c}}^{2}}   \exp\left(\frac{U_0}{T_H}\right) \right] \right. \\
 \left. +2\alpha\varphi_{0}\varphi_{1}\tilde{\varphi_{1}}{\varphi_{c}}^{2}\exp(U_{0}/T_{H}) +\varphi_{0}{\varphi_{1}}^{2}{\varphi_{c}}^{2}\frac{\alpha T_H}{2 U_0} \left(exp(2U_{0}/T_{H})-1 \right) \right)
 \end{split}
\end{equation}
\begin{equation}
\begin{split}
num2=\frac{{{\mu_{0}}^{2}}\lambda_{0}(1-\alpha)T_{C}}  {2U_0} \left[ \exp \left( \frac{2U_0}{T_C} \right) -1\right] +\frac{\xi_{1}(1-\alpha)T_C}{U_0} \left[ \exp \left(\frac{U_0}{T_C} \right) -1\right] +\xi_{0}(1-\alpha) \\
+\mu_{2}\lambda_{1}\frac{\alpha-1}{U_0}T_{C}\left[\exp\left( -\frac{U_0}{T_C} \right)-1 \right]
\end{split}
\end{equation}
\end{widetext}

The numerator is then given by $num=num1+num2$. The effective diffusion coefficient then obtained as
\begin{equation}
 D_{eff}=\frac{num}{{I_d}^3}
\end{equation}
The various terms are provided in the Appendix.
The current is  calculated as,
\begin{equation}
 <\dot{x}>=L\frac{1-\exp(\psi(L))}{I_d}
\end{equation}

In the low temperature limit, the particle current and effective diffusion coefficient
can be obtained in terms of the transition rates in forward and reverse directions given
by,
\begin{equation}
 r_{f}=\frac{1}{\alpha}\frac{1}{\alpha T_{H} Y^2 +(\alpha-1)T_{C}Z}
\end{equation}
where, 
\begin{equation}
 Y=\exp(\frac{U_{0}}{2T_{H}})-\exp(\frac{U_{0}}{2T_{C}})
\end{equation}
and,
\begin{equation}
 Z=(\exp(\frac{-U_{0}}{T_{C}})-1)(\exp(\frac{U_{0}}{T_{H}})-1)
\end{equation}
and,
\begin{equation}
 r_{b}=\frac{1}{1-\alpha}\frac{1}{\alpha T_{H} X +(1-\alpha) T_{C} R^{2}}
\end{equation}
where,
\begin{equation}
 X=(1-\exp(-\frac{1}{T_{H}}))(\exp(\frac{1}{T_{C}})-1)
\end{equation}
and,
\begin{equation}
 R=\exp(\frac{U_{0}}{2T_{C}})-\exp(-\frac{U_{0}}{2T_{C}})
\end{equation}
So, the current and effective diffusion coefficient are given by,
\begin{equation}
 <\dot{x}>=r_{f}\alpha-r_{b}(1-\alpha)
\end{equation}
and,
\begin{equation}
 D_{eff}=\frac{r_{f}\alpha+r_{b}(1-\alpha)}{2}
\end{equation}

{\it{\underline{Results.-}}} In Fig.~2, we plot the current as a function of the asymmetry parameter.
In FIg.~3 we show the effective diffusion coefficient as a function of the temperature 
of the hot bath. Fig.~4 shows the Peclet number. Good agreement is obtained between theory
and simulation result.

\begin{figure}[htb]
\includegraphics[width=3.0in]{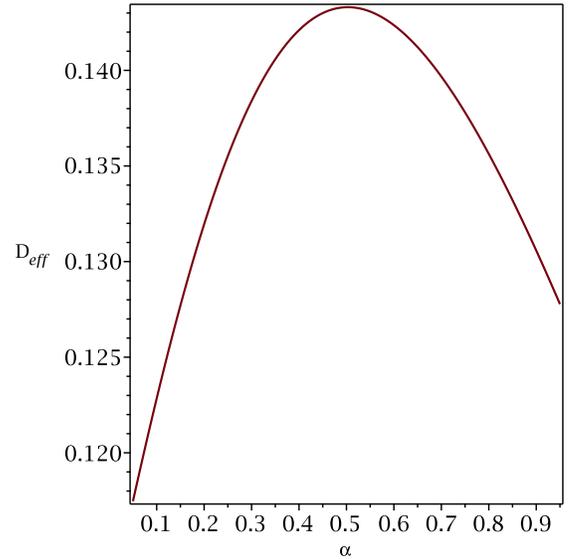}
\caption{\label{fig:Q-F_fridge}(Color online) Effective diffusion coefficient as a function
of the asymmetry parameter $\alpha$.}
\end{figure}

\begin{figure}[htb]
\includegraphics[width=3.0in]{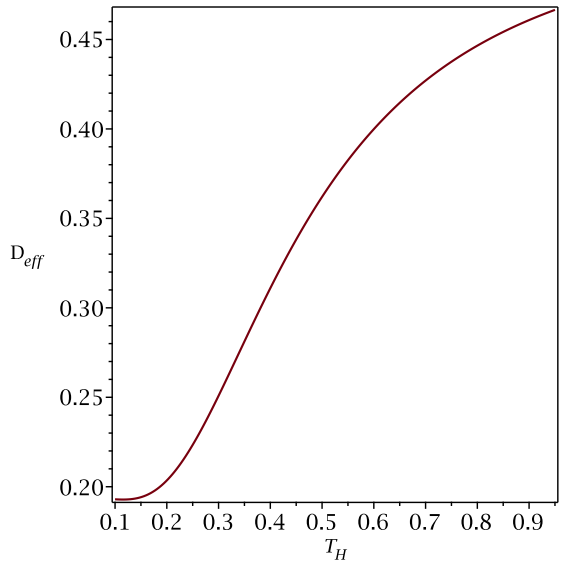}
\caption{\label{fig:Q-F_fridge}(Color online) Effective diffusion coefficient as a function
of the temperature of the hot bath.}
\end{figure}

\begin{figure}[htb]
\includegraphics[width=3.0in]{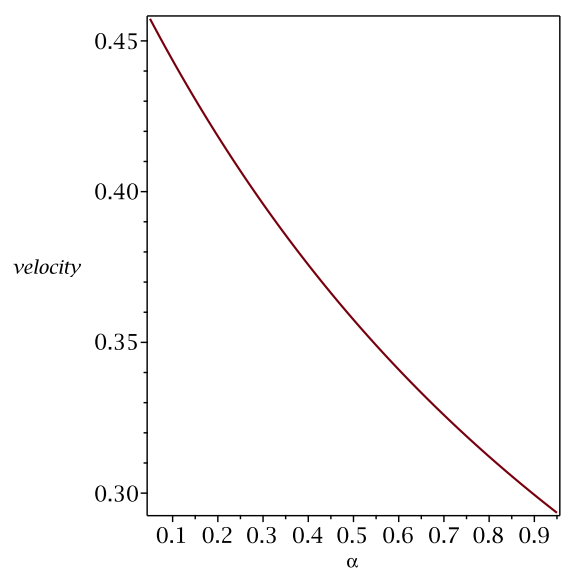}
\caption{\label{fig:Q-F_fridge}(Color online) Current as a function
of the asymmetry parameter $\alpha$.}
\end{figure}

\begin{figure}[htb]
\includegraphics[width=3.0in]{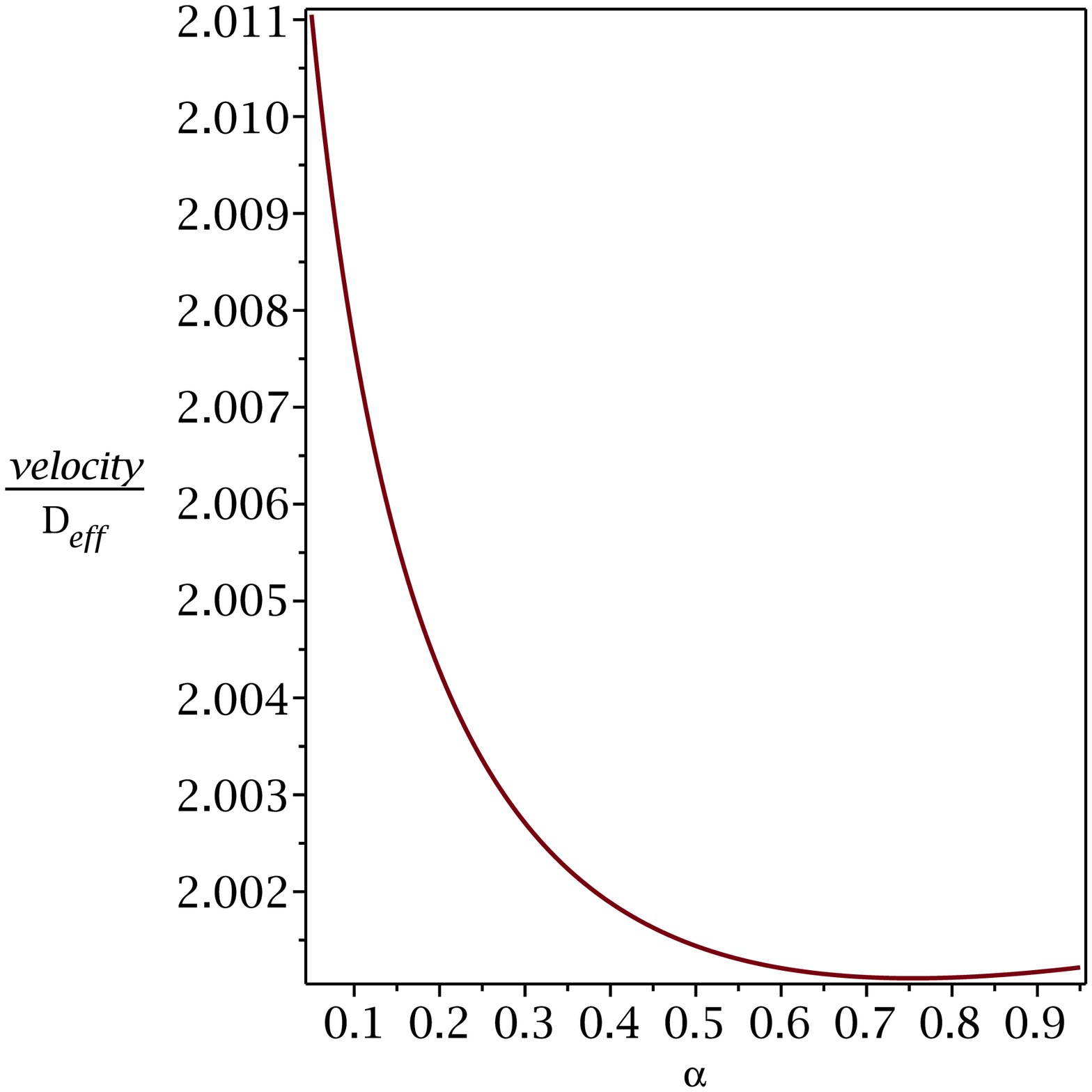}
\caption{\label{fig:Q-F_fridge}(Color online) Peclet number as a function
of the asymmetry parameter $\alpha$.}
\end{figure}

{\it{\underline{Conclusion.-}}} In this work, we have analytically and numerically obtained
the current and effective diffusion coefficient of a Brownian particle in a piecewise
linear potential subject alternately to hot and cold baths. Good agreement is obtained between
our numerical and analytical results. In some parameter regimes transport is enhanced.

\begin{acknowledgments}
The author acknowledges 
\end{acknowledgments}


\begin{thebibliography}{0}
\expandafter\ifx\csname natexlab\endcsname\relax\def\natexlab#1{#1}\fi
\expandafter\ifx\csname bibnamefont\endcsname\relax
  \def\bibnamefont#1{#1}\fi
\expandafter\ifx\csname bibfnamefont\endcsname\relax
  \def\bibfnamefont#1{#1}\fi
\expandafter\ifx\csname citenamefont\endcsname\relax
  \def\citenamefont#1{#1}\fi
\expandafter\ifx\csname url\endcsname\relax
  \def\url#1{\texttt{#1}}\fi
\expandafter\ifx\csname urlprefix\endcsname\relax\def\urlprefix{URL }\fi
\providecommand{\bibinfo}[2]{#2}
\providecommand{\eprint}[2][]{\url{#2}}

\end{thebibliography}


\begin{thebibliography}{99}
\bibitem{sekimoto97}
K.~Sekimoto, J. Phys. Soc. Jpn. {\bf 66},  1234 (1997).

\bibitem{sekimoto98}
K.~Sekimoto, Prog. Theor. Phys. Suppl. {\bf 130}, 17 (1998).

\bibitem{sekimoto}
K.~Sekimoto, \emph{Stochastic Energetics} (Springer, in preparation).

\bibitem{reimann02}
P.~Reimann, Phys. Rep. {\bf 361}, 57 (2002).

\bibitem{feynman}
R.~P.~Feynman, R.~B.~Leighton, and M.~Sands, \emph{The Feynman Lectures
on Physics} (Addison Wesley, Reading, MA, 1966), Vol. I, Chap. 46.

\bibitem{buttiker87}
M.~B{\"u}ttiker, Z. Phys. B {\bf 68}, 161 (1987).

\bibitem{landauer88}
R.~Landauer, J. Stat. Phys. {\bf 53}, 233 (1988).

\bibitem{parrondo96}
J.~M.~R.~Parrondo and P.~Espa{\~n}ol, Am. J. Phys. {\bf 64}, 1125
(1996).

\bibitem{hondou98}
T.~Hondou and F.~Takagi, J. Phys. Soc. Jpn. {\bf 67}, 2974 (1998).

\bibitem{vandenbroeck04}
C.~Van den Broeck, R.~Kawai, and P.~Meurs, Phys. Rev. Lett. {\bf 93},
090601 (2004).

\bibitem{kestemont00} 
E.~Kestemont, C.~Van den Broeck, and M.~Malek Mansour, Europhys. Lett.
{\bf 49}, 143 (2000).

\bibitem{vandenbroeck01}
C.~Van den Broeck, E.~Kestemont, and M.~Malek Mansour, Europhys. Lett.
{\bf 56}, 771 (2001).

\bibitem{jarzynski99}
C.~Jarzynski and O.~Mazonka, Phys. Rev. E {\bf 59}, 6448 (1999).

\bibitem{vandenbroeck06}
C.~Van den Broeck and R.~Kawai, Phys. Rev. Lett. {\bf 96}, 210601 (2006)

\bibitem{nakagawa06}
N.~Nakagawa and T.~S.~Komatsu, Europhys. Lett. {\bf 75}, 22 (2006)

\bibitem{bier96}
M.~Bier and R.~D.~Astumian, Bioelectrochem. Bioenerg. {\bf 39}, 67
(1996).

\bibitem{blanter98}
Y.~M.~Blanter and M.~B{\"u}ttiker, Phys. Rev. Lett. {\bf 81}, 4040
(1998).

\bibitem{derenyi99}
I.~Der\'{e}nyi and R.~D.~Astumian, Phys. Rev. E {\bf 59}, R6219 (1999).

\bibitem{hondou00}
T.~Hondou and K.~Sekimoto, Phys. Rev. E {\bf 62}, 6021 (2000).

\bibitem{matsuo00}
M.~Matsuo and S.~-I.~Sasa, Physica A {\bf 276}, 188 (2000)

\bibitem{asfaw04+05+07}
M.~Asfaw and M.~Bekele, Eur. Phys. J. B {\bf 38},  457 (2004);
Phys. Rev. E {\bf 72}, 056109 (2005);
Physica A {\bf 384}, 346 (2007).

\bibitem{ai05+06}
B.~-Q.~Ai, H.~-Z.~Xie, D.~-H.~Wen, X.~-M.~Liu, and L.~-G.~Liu, Eur.
Phys. J. B
{\bf 48}, 101 (2005);
B.~-Q.~Ai, L.~Wang, and L.~-G.~Liu, Phys. Lett. A {\bf 352}, 286
(2006).

\bibitem{vankampen91}
N.~G.~van Kampen, J. Stat. Phys. {\bf 63}, 1019 (1991).

\bibitem{vankampen88}
N.~G.~van Kampen, IBM J. Res. Dev.  {\bf 32}, 107 (1988).

\bibitem{boundary}
The boundary conditions for a system with inhomogeneous temperature is
discussed in Ref.~\cite{landauer88}. Different boundary conditions,
namely $P_{1}(0)=P_{2}(L)$ and $P_{1}(L/2)=P_{2}(L/2)$ are often used in
the literature \cite{bier96,asfaw04+05+07}. However, these boundary
conditions are not consistent with the physical system under
consideration. Indeed, the solution obtained with these boundary
conditions  disagrees with the results of our molecular dynamics
simulation both qualitatively and quantatively.

\bibitem{sancho92}
J.~M.~Sancho, M.~S.~Miguel, and D.~Duerr, J. Stat. Phys. {\bf 28}, 291
(1982).

\bibitem{jayannavar95}
A.~M.~Jayannavar and M.~C.~Mahato, Pramana J. Phys. {\bf 45}, 369
(1995).

\bibitem{meurs04}
P.~Meurs, C.~Van den Broeck, and A.~Garcia, Phys. Rev. E {\bf 70},
051109 (2004).

\bibitem{degroot}
S.~R.~de Groot and P.~Mazur, \emph{Non-Equilibrium Thermodynamics}
(Dover, New York, 1984).

\bibitem{vandenbroeck07}
C.~Van den Broeck, Adv. Chem. Phys. {\bf 135},  189 (2007).


\end{thebibliography}
\end{document}